\documentclass[a4paper,12pt]{article}
\usepackage{jheppub} % for details on the use of the package, please see the JINST-author-manual
\usepackage{lineno}
\usepackage{bbold, mathrsfs, tikz, amssymb, simpler-wick, slashed, graphicx, dsfont}
\usepackage{eucal}
\usepackage[export]{adjustbox}
\usepackage[italicdiff]{physics}
\newcommand{\ieps}[0]{i \epsilon}

\newcommand{\amp}[0]{\mathcal{A}}
\newcommand{\reals}[0]{\mathds{R}}

\newcommand{\nn}[0]{\nonumber \\}
\newcommand{\qandq}[0]{\quad \text{and} \quad}
\newcommand{\qwhere}[0]{\quad \text{where} \quad}
\newcommand{\npar}[0]{\bigskip \par \noindent}

\newcommand\be{\begin{equation}}
\newcommand\ba{\begin{eqnarray}}
\newcommand\ee{\end{equation}}
\newcommand\ea{\end{eqnarray}}
\newcommand\bw{\begin{widetext}}
\newcommand\ew{\end{widetext}}
\usepackage[normalem]{ulem}

\tikzstyle{block} = [draw,rectangle,thick,minimum height=2em,minimum width=2em]
\tikzstyle{sum} = [draw,circle,inner sep=0mm,minimum size=2mm]
\tikzstyle{connector} = [->,thick]
\tikzstyle{line} = [thick]
\tikzstyle{branch} = [circle,inner sep=0pt,minimum size=1mm,fill=black,draw=black]
\tikzstyle{guide} = []
\tikzstyle{snakeline} = [connector, decorate, decoration={pre length=0.2cm,
                         post length=0.2cm, snake, amplitude=.4mm,
                         segment length=2mm},thick, blue, ->]
\tikzset{graviton/.style={decorate, decoration={snake, amplitude=.3mm, segment length=1.3mm, pre length=0mm, post length=0mm}, double}}
\usepackage[compat=1.1.0]{tikz-feynman}
\usepackage[compat=1.1.0]{tikz-feynhand}
\allowdisplaybreaks

\title{\boldmath{Worldline Proof of Eikonal Exponentiation}}

\author[a, b]{Yuchen Du,}
\author[a]{Siddarth Ajith,}
\author[a]{Ravisankar Rajagopal,}
\author[a]{and Diana Vaman}
\affiliation[a]{Department of Physics,  University of Virginia, \\ Charlottesville, Virginia 22904-4714, USA}
\affiliation[b]{School of Physics and Shing-Tung Yau Center, Southeast University, \\
Nanjing 210018, China}

\emailAdd{sa4fb@virginia.edu, yd3gh@seu.edu.cn, rr2yj@virginia.edu, dv3h@virginia.edu}

\abstract{In this paper, working in the eikonal approximation, we present a proof for the exponentiation of the 2-body eikonal phase  to {\it all orders in the eikonal expansion},  for scalar particles interacting electromagnetically or gravitationally.  The proof is based on the worldline formalism, which is an alternative, first quantized method to the standard QFT calculation of the scattering amplitude.  We show that in the worldline formalism the 2-body scattering amplitude written in impact parameter space naturally factorizes at each loop order. This factorization is responsible for the exponentiation of the eikonal phase, a result which was anticipated in the work of Mogull, Plefka, and Steinhoff \cite{Mogull_2021}.}

\begin{document}
\maketitle
\flushbottom

\input{wl.tex}

\appendix
\section{Examples of Factorization in $2\to2$ Scattering}\label{gravity-example}
Here, we present two simple examples to help understand the factorization of
$2 \rightarrow 2$ scattering in general cases. In the first example, we focus on explaining the {bookkeeping devices}
$t_{mn}$ which count how many contractions were made between the connected mediator ``$m$" and ``$n$" subdiagrams, and the reason why we {need them}. In the second
example, we will explicitly show how the procedure works.

\subsubsection*{First Example}
Let us consider the following two diagrams
which appear at the same $\hbar$ order. \be
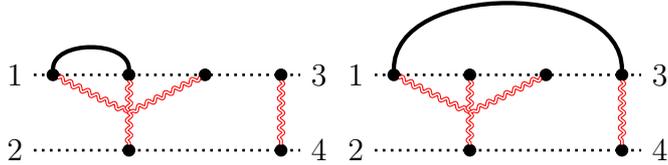
\begin{tikzpicture}[baseline=0.4cm]
   \begin{feynhand}
      \vertex (b) at (-1.5,0){2}; 
      \vertex (d) at (2.5,0){4};
      \vertex (a) at (-1.5,1){1}; 
      \vertex (c) at (2.5,1){3};
      \propag [gho] (a) to (c); 
      \propag [gho] (b) to (d);
      
      \vertex [dot](t1t3) at (1,1){};
      \vertex [dot](t1t2) at (0,1){};
      \vertex [dot](t1t1) at (-1,1){};
      \vertex [dot](t1b1) at (0,0){};
      \vertex [dot](t1m1) at (0,0.5);
      \propag [graviton,red] (t1t1) to (t1m1);
      \propag [graviton,red] (t1t2) to (t1m1);
      \propag [graviton,red] (t1t3) to (t1m2);
      \propag [graviton,red] (t1m1) to (t1b1);

      \vertex [dot](t2t1) at (2, 1){};
      \vertex [dot](t2b1) at (2, 0){};
      \propag [graviton,red] (t2t1) to (t2b1);

      %%%%%%% contraction zone %%%%%%%%%%%%%%%
      \propag [plain, ultra thick] (t1t1) [in=90, out=90, looseness=1] to (t1t2);
   \end{feynhand}
\end{tikzpicture}
\begin{tikzpicture}[baseline=0.4cm]
   \begin{feynhand}
      \vertex (b) at (-1.5,0){2}; 
      \vertex (d) at (2.5,0){4};
      \vertex (a) at (-1.5,1){1}; 
      \vertex (c) at (2.5,1){3};
      \propag [gho] (a) to (c); 
      \propag [gho] (b) to (d);
      
      \vertex [dot](t1t3) at (1,1){};
      \vertex [dot](t1t2) at (0,1){};
      \vertex [dot](t1t1) at (-1,1){};
      \vertex [dot](t1b1) at (0,0){};
      \vertex [dot](t1m1) at (0,0.5);
      \propag [graviton,red] (t1t1) to (t1m1);
      \propag [graviton,red] (t1t2) to (t1m1);
      \propag [graviton,red] (t1t3) to (t1m2);
      \propag [graviton,red] (t1m1) to (t1b1);

      \vertex [dot](t2t1) at (2, 1){};
      \vertex [dot](t2b1) at (2, 0){};
      \propag [graviton,red] (t2t1) to (t2b1);

      %%%%%%% contraction zone %%%%%%%%%%%%%%%
      \propag [plain, ultra thick] (t1t1) [in=90, out=90, looseness=1] to (t2t1);
   \end{feynhand}
\end{tikzpicture} \nn
\ee
They cover all possible choices of forming one contraction on the first
worldline. It is obvious to see from the expression $\langle \mathcal{T} [V_1
V_2 V_3 V_4] \rangle$ that the total number of the choices is $\binom{4}{2} =
6$. More specifically, 3 choices for forming the contraction in the first
diagram and 3 choices for forming the contraction in the second diagram. 

\npar However, in utilizing the  $t_{mn}$ to do the counting,  one only distinguishes between the contractions between tree
parts. In this case we only have two connected mediator subdiagrams, one ``$\Psi$" tree and one ``I" tree.
The first diagram is simply represented by $\tilde A|_1$ (or $\tilde A|_{t_{12}^0}$) since there is no contraction between the two trees, while the second diagram is represented by $\tilde A|_{t_{12}}$. {From the point view of counting diagrams  using $t_{mn}$, we only have two choices, either performing the contraction between the two trees or not.}
\npar So how do we reconcile the two countings, 6 choices {in total} and 2 choices when using $t_{mn}$? We can think of the whole procedure of counting as done in 
two steps. The first step is to count how many ways one can distribute
contractions into separated groups of tree parts, such that tree parts in each
group are connected at least. And the second step is to count how many ways
one can perform contractions within each group, under the condition that tree
parts are connected. 
Thus the
counting in this example can be understood as
\begin{eqnarray*}
  &  & 1 \left( \text{ways to form 1 group of connected tree parts} \right)
  \times 3 \left( \text{ways of doing contractions in 1st group} \right)\\
  & + & 1 \left( \text{ways to form 2 groups of connected tree parts} \right)
  \times 3 \left( \text{ways in 1st group} \right) \times 1 \left( \text{ways
  in 2nd group} \right)
\end{eqnarray*}
Now let us explain why the counting of $t_{mn}$ matters by focusing on the first diagram,
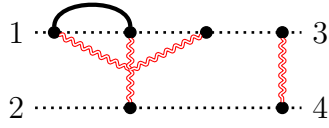
\be
\begin{tikzpicture}[baseline=0.4cm]
   \begin{feynhand}
      \vertex (b) at (-1.5,0){2}; 
      \vertex (d) at (2.5,0){4};
      \vertex (a) at (-1.5,1){1}; 
      \vertex (c) at (2.5,1){3};
      \propag [gho] (a) to (c); 
      \propag [gho] (b) to (d);
      
      \vertex [dot](t1t3) at (1,1){};
      \vertex [dot](t1t2) at (0,1){};
      \vertex [dot](t1t1) at (-1,1){};
      \vertex [dot](t1b1) at (0,0){};
      \vertex [dot](t1m1) at (0,0.5);
      \propag [graviton,red] (t1t1) to (t1m1);
      \propag [graviton,red] (t1t2) to (t1m1);
      \propag [graviton,red] (t1t3) to (t1m2);
      \propag [graviton,red] (t1m1) to (t1b1);

      \vertex [dot](t2t1) at (2, 1){};
      \vertex [dot](t2b1) at (2, 0){};
      \propag [graviton,red] (t2t1) to (t2b1);

      %%%%%%% contraction zone %%%%%%%%%%%%%%%
      \propag [plain, ultra thick] (t1t1) [in=90, out=90, looseness=1] to (t1t2);
   \end{feynhand}
\end{tikzpicture} \nn
\ee
This diagram is represented by $\eval{\widetilde{\mathcal{A}}}_1$. As a
consequence of the factorization of the worldline amplitude,  $\eval{\tilde{A}}_1$ can be factorized and we have
\[ \eval{\widetilde{\mathcal{A}}_3}_1 = x \eval{\widetilde{\mathcal{A}}_2}_1 \times \eval{\widetilde{\mathcal{A}}_0}_1, \]
where $x$ is a  coefficient to be determined, and $\eval{\widetilde{\mathcal{A}}_1}_1$ and $\eval{\widetilde{\mathcal{A}}_2}_1$ represent the diagrams
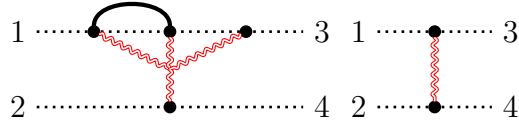
\be
\begin{tikzpicture}[baseline=0.4cm]
   \begin{feynhand}
      \vertex (b) at (-2,0){2}; 
      \vertex (d) at (2,0){4};
      \vertex (a) at (-2,1){1}; 
      \vertex (c) at (2,1){3};
      \propag [gho] (a) to (c); 
      \propag [gho] (b) to (d);
      
      \vertex [dot](t1t3) at (1,1){};
      \vertex [dot](t1t2) at (0,1){};
      \vertex [dot](t1t1) at (-1,1){};
      \vertex [dot](t1b1) at (0,0){};
      \vertex [dot](t1m1) at (0,0.5);
      \propag [graviton,red] (t1t1) to (t1m1);
      \propag [graviton,red] (t1t2) to (t1m1);
      \propag [graviton,red] (t1t3) to (t1m2);
      \propag [graviton,red] (t1m1) to (t1b1);

      %%%%%%% contraction zone %%%%%%%%%%%%%%%
      \propag [plain, ultra thick] (t1t1) [in=90, out=90, looseness=1] to (t1t2);
   \end{feynhand}
\end{tikzpicture}
\begin{tikzpicture}[baseline=0.4cm]
   \begin{feynhand}
      \vertex (b) at (1,0){2}; 
      \vertex (d) at (3,0){4};
      \vertex (a) at (1,1){1}; 
      \vertex (c) at (3,1){3};
      \propag [gho] (a) to (c); 
      \propag [gho] (b) to (d);

      \vertex [dot](t2t1) at (2, 1){};
      \vertex [dot](t2b1) at (2, 0){};
      \propag [graviton,red] (t2t1) to (t2b1);

      %%%%%%% contraction zone %%%%%%%%%%%%%%%
   \end{feynhand}
\end{tikzpicture} \nn
\ee
Solving for $x$, we have
\[ x = \frac{\eval{\widetilde{\mathcal{A}}_3}_1}{\eval{\widetilde{\mathcal{A}}_2}_1 \times
\eval{\widetilde{\mathcal{A}}_0}_1} \]
After the contractions and integrals cancelled out between the numerator and the denominator, this becomes
\begin{eqnarray*}
  x & = & \frac{\# \times \left(
  \text{ways to form 2 groups} \right) \times \left( \text{ways in 1st group}
  \right) \times \left( \text{ways in 2nd group} \right)}{\#_1 \times \left( \text{ways in 1st group} \right) \times
  \#_2 \times \left( \text{ways in 2nd group} \right)}\\
  & = & \frac{\# \times \left( \text{ways to form 2 groups}
  \right)}{\#_1 \times \#_2}
\end{eqnarray*}
where $\#_1, \#_2$ and $\#$ are the symmetry factors from the worldline expression of scattering amplitude in \eqref{eq-sym-factor}. Notice that the numbers of ways to do the contractions within
each group are simply cancelled out. Thus, only the counting from $t_{mn}$ (in this case it is the "ways to form 2 groups")
really matters.

\subsubsection*{Second Example}
Let us consider the following diagram which has five connected graviton $n$-point tree diagrams: 
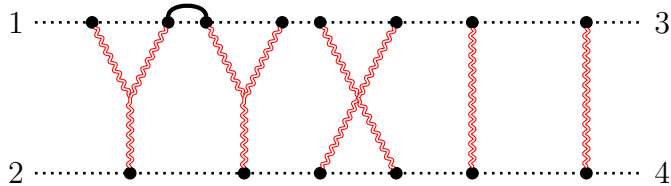
\be
\begin{tikzpicture}[baseline=1cm]
   \begin{feynhand}
      \vertex (b) at (-4.5,0){2}; 
      \vertex (d) at (4,0){4};
      \vertex (a) at (-4.5,2){1}; 
      \vertex (c) at (4,2){3};
      \propag [gho] (a) to (c); 
      \propag [gho] (b) to (d);
      
      \vertex [dot](t1t1) at (-3.5, 2){};
      \vertex [dot](t1t2) at (-2.5, 2){};
      \vertex [dot](t1m1) at (-3, 1);
      \vertex [dot](t1b1) at (-3, 0){};
      \propag [graviton,red] (t1t1) to (t1m1);
      \propag [graviton,red] (t1t2) to (t1m1);
      \propag [graviton,red] (t1b1) to (t1m1);

      \vertex [dot](t2t1) at (-2, 2){};
      \vertex [dot](t2t2) at (-1, 2){};
      \vertex [dot](t2m1) at (-1.5, 1);
      \vertex [dot](t2b1) at (-1.5, 0){};
      \propag [graviton,red] (t2t1) to (t2m1);
      \propag [graviton,red] (t2t2) to (t2m1);
      \propag [graviton,red] (t2b1) to (t2m1);

      \vertex [dot](t3t1) at (-0.5, 2){};
      \vertex [dot](t3t2) at (0.5, 2){};
      \vertex [dot](t3m1) at (0, 1);
      \vertex [dot](t3b1) at (-0.5, 0){};
      \vertex [dot](t3b2) at (0.5, 0){};
      \propag [graviton,red] (t3t1) to (t3m1);
      \propag [graviton,red] (t3t2) to (t3m1);
      \propag [graviton,red] (t3b1) to (t3m1);
      \propag [graviton,red] (t3b2) to (t3m1);

      \vertex [dot](t4t1) at (1.5, 2){};
      \vertex [dot](t4b1) at (1.5, 0){};
      \propag [graviton,red] (t4t1) to (t4b1);

      \vertex [dot](t5t1) at (3, 2){};
      \vertex [dot](t5b1) at (3, 0){};
      \propag [graviton,red] (t5t1) to (t5b1);
      
      %%%% contraction zone %%%%
      \propag [plain, ultra thick] (t1t2) [in=90, out=90, looseness=1] to (t2t1);
   \end{feynhand}
\end{tikzpicture} \nn
\ee
\npar The diagram is represented by $\eval{\widetilde{\mathcal{A}}_8}_{t_{12}}$, where $t_{12}$ denotes the link between the first two graviton trees. We have two ``Y" tree parts, one ``X" tree part and two
``I" tree parts. Given our definitions for  $n$, the number of distinct kinds of tree
parts, $c_i$, the number of tree parts in each kind, $u_i$ and $d_i$, the number
of legs attached to the top and bottom  worldlines for each of the graviton trees, 
we have
\begin{eqnarray*}
  &  & n = 3\\
  \text{``Y"} &  & c_1 = 2, u_1 = 2, d_1 = 1\\
  \text{``X"} &  & c_2 = 1, u_2 = 2, d_2 = 2\\
  \text{``I"} &  & c_3 = 2, u_3 = 1, d_3 = 1\,.
\end{eqnarray*}
Next we want to put them into the following four separate groups,
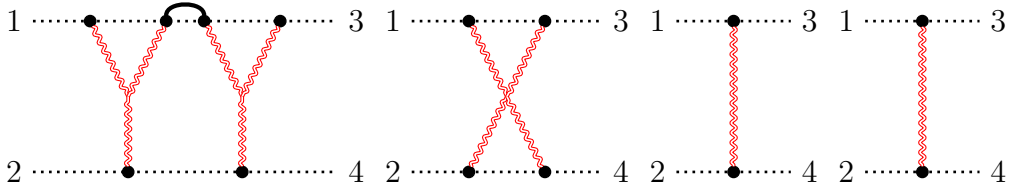
\be
\begin{tikzpicture}[baseline=1cm]
   \begin{feynhand}
      \vertex (b) at (-4.5,0){2}; 
      \vertex (d) at (0,0){4};
      \vertex (a) at (-4.5,2){1}; 
      \vertex (c) at (0,2){3};
      \propag [gho] (a) to (c); 
      \propag [gho] (b) to (d);
      
      \vertex [dot](t1t1) at (-3.5, 2){};
      \vertex [dot](t1t2) at (-2.5, 2){};
      \vertex [dot](t1m1) at (-3, 1);
      \vertex [dot](t1b1) at (-3, 0){};
      \propag [graviton,red] (t1t1) to (t1m1);
      \propag [graviton,red] (t1t2) to (t1m1);
      \propag [graviton,red] (t1b1) to (t1m1);

      \vertex [dot](t2t1) at (-2, 2){};
      \vertex [dot](t2t2) at (-1, 2){};
      \vertex [dot](t2m1) at (-1.5, 1);
      \vertex [dot](t2b1) at (-1.5, 0){};
      \propag [graviton,red] (t2t1) to (t2m1);
      \propag [graviton,red] (t2t2) to (t2m1);
      \propag [graviton,red] (t2b1) to (t2m1);
      
      %%%% contraction zone %%%%
      \propag [plain, ultra thick] (t1t2) [in=90, out=90, looseness=1] to (t2t1);
   \end{feynhand}
\end{tikzpicture}
\begin{tikzpicture}[baseline=1cm]
   \begin{feynhand}
      \vertex (b) at (-1.5,0){2}; 
      \vertex (d) at (1.5,0){4};
      \vertex (a) at (-1.5,2){1}; 
      \vertex (c) at (1.5,2){3};
      \propag [gho] (a) to (c); 
      \propag [gho] (b) to (d);

      \vertex [dot](t3t1) at (-0.5, 2){};
      \vertex [dot](t3t2) at (0.5, 2){};
      \vertex [dot](t3m1) at (0, 1);
      \vertex [dot](t3b1) at (-0.5, 0){};
      \vertex [dot](t3b2) at (0.5, 0){};
      \propag [graviton,red] (t3t1) to (t3m1);
      \propag [graviton,red] (t3t2) to (t3m1);
      \propag [graviton,red] (t3b1) to (t3m1);
      \propag [graviton,red] (t3b2) to (t3m1);
      
      %%%% contraction zone %%%%

   \end{feynhand}
\end{tikzpicture}
\begin{tikzpicture}[baseline=1cm]
   \begin{feynhand}
      \vertex (b) at (0.5,0){2}; 
      \vertex (d) at (2.5,0){4};
      \vertex (a) at (0.5,2){1}; 
      \vertex (c) at (2.5,2){3};
      \propag [gho] (a) to (c); 
      \propag [gho] (b) to (d);

      \vertex [dot](t4t1) at (1.5, 2){};
      \vertex [dot](t4b1) at (1.5, 0){};
      \propag [graviton,red] (t4t1) to (t4b1);
      
      %%%% contraction zone %%%%
   \end{feynhand}
\end{tikzpicture}
\begin{tikzpicture}[baseline=1cm]
   \begin{feynhand}
      \vertex (b) at (2,0){2}; 
      \vertex (d) at (4,0){4};
      \vertex (a) at (2,2){1}; 
      \vertex (c) at (4,2){3};
      \propag [gho] (a) to (c); 
      \propag [gho] (b) to (d);

      \vertex [dot](t5t1) at (3, 2){};
      \vertex [dot](t5b1) at (3, 0){};
      \propag [graviton,red] (t5t1) to (t5b1);
      
      %%%% contraction zone %%%%
   \end{feynhand}
\end{tikzpicture}
\nn
\ee
\npar Notice that the last two groups are indistinguishable. But let us first assume
that they are all distinguishable, and count how many ways we can fill in those
groups with our 5 graviton connected subdiagrams (which in this example are all trees, and we will refer to as graviton trees). We do this counting in a systematic way that can
be easily generalized to the most general cases.

\npar First, to fill in the first group, we need to take 2 ``Y" from all the ``Y"
graviton trees, 0 ``X" from all the ``X" trees, and 0 ``I" from all the
``I" trees, which is
\[ \binom{2}{2} \binom{1}{0} \binom{2}{0} = 1. \]
Next, to fill in the second group, we need to take 0 ``Y", 1 ``X" and 0 ``I" from
all trees left after the previous round of picking,
\[ \binom{2 - 2}{0} \binom{1}{1} \binom{2}{0} = 1. \]
Then, we fill in the third group,
\[ \binom{2 - 2}{0} \binom{1 - 1}{0} \binom{2}{1} = 2. \]
At last, we fill in the last group,
\[ \binom{2 - 2}{0} \binom{1 - 1}{0} \binom{2 - 1}{1} = 1 \]
In total, we have $1 \times 1 \times 2 \times 1 = 2$ ways of filling in the four
separate groups. However, since the last two groups are actually
indistinguishable, we need to divide this number by $2!$ to get the correct
counting, which is simply 1.
We can verify this number by simply looking at the expression of the first
worldline,
\[ \langle \mathcal{T} [(V_1 V_2)^{(1)} (V_3 V_4)^{(2)} (V_5)^{(3)} 
   (V_6)^{(4)}  (V_7)^{(5)}] \rangle, \]
where we use $(1), (2) \ldots$ to mark the tree parts that those vertex
operators belong to. There is only contraction to be done between the graviton trees
$(1)$ and $(2)$ to form group 1. Then obviously we only have 1 choice to put
these trees into the corresponding 4 groups! This is just another way to
say the expression directly factorizes,
\[ \langle \mathcal{T} [(V_1 V_2)^{(1)} (V_3 V_4)^{(2)}] \rangle \times \langle
   (V_5)^{(3)} \rangle \times \langle (V_6)^{(4)} \rangle \times \langle
   (V_7)^{(5)} \rangle . \]

\npar As a consequence of the factorization of the worldline amplitude, 
$\eval{\widetilde{\mathcal{A}}_8}_1$ can be factorized
\begin{eqnarray*}
  \eval{\widetilde{\mathcal{A}}_8}_{t_{12}} & = & x
  \eval{\widetilde{\mathcal{A}}_3}_{t_{12}} \times
  \eval{\widetilde{\mathcal{A}}_2}_1 \times \eval{\widetilde{\mathcal{A}}_0}_1 \times \eval{\widetilde{\mathcal{A}}_0}_1\,.
\end{eqnarray*}
Notice that the diagram of the third term $\widetilde{\mathcal{A}}_0$ is actually the same as the one of the fourth term
$\widetilde{\mathcal{A}}_0$\footnote{As we mentioned in section \ref{ssec:fac-2to2-general}, $\widetilde{\mathcal{A}}_L$ does not represent the full amplitude. Thus, the two $\widetilde{\mathcal{A}}_0$ could be different in principle. Whether they are the same or not depends on the diagrams they represent.}, the factorization is
\[ \begin{array}{lll}
     \eval{\widetilde{\mathcal{A}}_8}_{t_{12}} & = & x
     \eval{\widetilde{\mathcal{A}}_3}_{t_{12}} \times \eval{\widetilde{\mathcal{A}}_2}_1
     \times (\eval{\widetilde{\mathcal{A}}_0}_1)^2
   \end{array} \]
To determine the $x$, since we only have 1 way to form the 4 groups, we have
\begin{eqnarray*}
  x & = & \frac{\#}{\#_1 \times \#_2 \times (\#_3)^2}
\end{eqnarray*}
From our $n$, $c_i$, $u_i$ and $d_i$, we have
\begin{eqnarray*}
  \#_1 & = & \frac{1}{2! \times (2!)^2 \times (1!)^2} = \frac{1}{8}\\
  \#_2 & = & \frac{1}{1! \times (2!)^1 \times (2!)^1} = \frac{1}{4}\\
  \#_3 & = & \frac{1}{1! \times (1!)^1 \times (1!)^1} = 1\\
  \# & = & \frac{1}{2! \times (2!)^2 \times (1!)^2} \times \frac{1}{1! \times
  (2!)^1 \times (2!)^1} \times \frac{1}{2! \times (1!)^1 \times (1!)^1} =
  \frac{1}{64}\\
  x & = & \frac{1}{2!}
\end{eqnarray*}
In other words, we have the factorization,
\[ \begin{array}{lll}
     \eval{\widetilde{\mathcal{A}}_8}_{t_{12}} & = & 
     \eval{\widetilde{\mathcal{A}}_3}_{t_{12}} \times
     \eval{\widetilde{\mathcal{A}}_2}_1 \times \frac{1}{2!}
     (\eval{\widetilde{\mathcal{A}}_0}_1)^2,
   \end{array} \]
which matches our general result
\be
\eval{\widetilde{\mathcal{A}}_L}_{P(\{t_{ij}\})} =
\qty[\prod_{j=1}^{n_G} \frac{1}{g_j!}]\prod_{k=1}^{N_G}\eval{\widetilde{A}_{L_k}}_{P_k(\{t_{i_k j_k}\})},
\ee
when total number of groups $N_G=4$, number of distinct groups $n_G=3$, with number of copies in each distinct kind $g_1=g_2=1$ and $g_3=2$
\section{$\hbar$ counting in 2-body gravitational interactions }\label{gravity-counting}
Let us consider the general 2-body scattering amplitude for scalars interacting
gravitationally \eqref{eq-amplitude-formula_general},
\begin{equation}
    \tilde{\amp}_L(b) = \# \int_{\bar k_1, \cdots, \bar k_N}^{} e^{i \bar{q}\cdot b} \tilde{\mathcal{M}}_{N_1}(p_1, p_1') \cdot \qty(T^{(1)}  T^{(2)} ...) \cdot \tilde{\mathcal{M}}_{N_2}(p_2, p_2')\,,
\end{equation}
where the worldline amplitude $\widetilde{\mathcal{M}}_N$ from \eqref{eq-wl-amplitude-with-extra-delta} is 
\begin{equation}
    \widetilde{\mathcal{M}}_N(p, p') = (\frac{i}{2} \kappa)^N \qty(\prod_{j=1}^{N} \int_{-\infty}^{\infty}d\tau_j e^{-\epsilon \abs*{\tau_j}})\expval{\mathcal{T}\qty(\hat V_1(\bar k_1, \tau_1) \hat V_2(\bar k_2, \tau_2) \cdots \hat V_N(\bar k_N, \tau_N))}.
\end{equation}
and the linear vertex operators were given in  \eqref{eq-vertex-gr},
\begin{equation}
    \hat{V_j}(\bar{k}_j, \tau_j) = (\epsilon_j)_{\mu\nu} (\bar{v}^{\mu}+\dot{\hat{x}}^{\mu}(\tau_i))(\bar{v}^{\nu}+\dot{\hat{x}}^{\nu}(\tau_i)) e^{i \bar{k}_j\cdot (\bar{v}\tau_j+\hat{x}(\tau_j))}.
\end{equation}
In principle, there could also be non-linear vertex operators coming from the 
non-minimal coupling $R\Phi^2$ between the scalar field and gravity. We will address this case later and first consider the case when the vertex operators are linear.

\npar Let us assume that we have a total number $N$ of graviton connected $n$-point  functions that attach to the two worldlines.  For each connected graviton  diagram, let us assume there are $n_i$ external legs and $l_i$ loops. Thus, the number of coupling constants coming with each $n$-point diagram is $(n_i + 2 l_i - 2)$.

\npar Let us first count the $\hbar$ factors for a WQFT diagram without any $x{-}x$  contractions. There are three sources for $\hbar$: the coupling constant $\kappa$ which contains $\hbar$, the factor $\left( {v_a^{\mu}}/{\hbar} \right) \left({v_a^{\nu}}/{\hbar} \right)$ in each vertex, and the exponential $e^{i\frac{v_a}{\hbar} \cdot k \tau}$ which will yield factors of $\hbar$ after performing the worldline time $\tau$ integral. The coupling constant $\kappa$ relates to the classical Newton's constant by $\kappa = \sqrt{8 \pi G \hbar}$, which can be derived by matching the QFT tree amplitude with the classical Newtonian potential.
%\npar The coupling constants can come from the worldline amplitude and each $n$-point parts. 
Counting all sources of $\hbar$ factors from the coupling constants we have
\begin{eqnarray*}
  \left( \hbar^{\frac{1}{2}} \right)^{\sum_{i = 1}^N (n_i + 2 l_i - 2) +
  \sum_{i = 1}^N n_i} & = & \hbar^{ \left( \sum_{i = 1}^N n_i \right) + l -
  N},
\end{eqnarray*}
where $l = \sum_{i = 1}^N l_i$ is the total number of loops in all the
$n$-point parts.

\npar Additionally, each vertex operator contributes to the leading order a factor $\left( \frac{v_a}{\hbar} \right)^{\mu}\left( \frac{v_a}{\hbar} \right)^{\nu}$. In all, they yield
\begin{eqnarray*}
  (\hbar^{- 2})^{\sum_i^N n_i} & = & {\hbar^{- 2 \sum_i^N n_i}}  .
\end{eqnarray*}
Each worldline time integral yields $\hbar$. In all, they yield
\[ \hbar^{ \sum_{i = 1}^N n_i} .\]
Thus, the $\hbar$ counting for a diagram without any contraction is
\begin{eqnarray}
  \hbar^{ \left( \sum_{i = 1}^N n_i \right) + l - N} {\times \hbar^{- 2
  \sum_i^N n_i}}  \times \hbar^{ \sum_{i = 1}^N n_i} & = & \hbar^{- N + l},
\end{eqnarray}
which is a rather simple result.

\npar Next, let us add contractions between the worldline vertices. As we discussed in Section \ref{ssec-hbar}, each contraction essentially adds one factor of $\hbar$. Assuming there are $n_c$ contractions, the final result of $\hbar$ counting for a diagram that has  only linear vertex operators is
\[ \hbar^{- N + l + n_c} . \]
Now, let us draw some conclusions from this counting, which will be used in the main body of the paper. Let us consider the case where $l = 0$, which means all $n$-point graviton connected diagrams  are trees. Since classical terms are of order $O
(\hbar^{- 1})$, to get a classical term, we need
\[ n_c = N - 1. \]
Recall that we exactly have $N$ tree parts, which require at least $N - 1$
contractions to make them connected through contractions.
In other words to form an irreducible diagram we need at least $N-1$ contractions. Thus, we arrived at our first conclusion. Diagrams with only graviton trees which are fully (i.e. irreducible) and minimally connected through $x{-}x$ contractions are classical. 

\npar A very useful case is the minimally connected ladder diagram, which can be used to compare with other diagrams to determine if they are superclassical, classical or quantum.
Notice that in reaching this conclusion, it does not matter what are the graviton trees. Thus, a diagram built with graviton trees will be of the  same $\hbar$ order as a ladder with the same number of coupling constants and which is (fully and) minimally connected. This observation  will be used in Section \ref{ssec-exponentiation-gr}.

\npar If $l\geq 1$, diagrams which have all graviton $n$-point functions  connected, which means $n_c \geqslant N - 1$,  will be of order  $\hbar^k$ with $k\geq 0$. Thus, we get our second conclusion: Diagram with induced loops in the $n$-point graviton diagrams which are fully connected through contractions, are quantum.

\npar These  two conclusions will be used in Section \ref{ssec-hbar-gr}.

\npar At last, let us address the issue of non-linear vertex operators. The non-minimal coupling corresponds to a term that is proportional to the Ricci scalar $R$ in the worldline action. When expanded in terms of the gravitational fields $h_{\mu \nu}$, it always contains two derivatives. However, since the wavenumber of the gravitons is taken to be of order $O(\hbar^0)$, a non-linear vertex by itself will only produce one factor of $\hbar$, which comes from the worldline time integral. In comparison, a {minimally connected ladder}, which has the same number of coupling constants, yields $\hbar^{-2}$ by itself. Thus, a connected diagram with non-linear vertex operators is always a quantum term.

\bibliographystyle{JHEP}
\bibliography{biblio.bib}

\end{document}